# Excitation properties of the divacancy in 4H-SiC

Björn Magnusson,[1,2] Nguyen Tien Son,[1] András Csóré,[3] Andreas Gällström,[4] Takeshi Ohshima,[5] Adam Gali,[3,6] Ivan G. Ivanov[1]

[1] Linköping University, Department of Physics, Chemistry and Biology, S-581 83 Linköping, Sweden
[2] Norstel AB, Ramshällsvägen 15, SE-602 38 Norrköping, Sweden
[3] Department of Atomic Physics, Budapest University of Technology and Economics, Budafoki út. 8, H-1111, Hungary
[4] Saab Dynamics AB, SE-581-88 Linköping, Sweden
[5] National Institutes for Quantum and Radiological Science and Technology, 1233 Watanuki, Takasaki, Gunma 370-1292, Japan
[6] Wigner Research Center for Physics, Hungarian Academy of Sciences, PO. Box 49, H-1525, Hungary

*Abstract*. We investigate the quenching of the photoluminescence (PL) from the divacancy defect in 4H-SiC consisting of a nearest-neighbour silicon and carbon vacancies. The quenching occurs only when the PL is excited below certain photon energies (thresholds), which differ for the four different inequivalent divacancy configurations in 4H-SiC. Accurate theoretical *ab initio* calculation for the charge-transfer levels of the divacancy show very good agreement between the position of the (0/−) level with respect to the conduction band for each divacancy configurations and the corresponding experimentally observed threshold, allowing us to associate the PL decay with conversion of the divacancy from neutral to negative charge state due to capture of electrons photoionized from other defects (traps) by the excitation. Electron paramagnetic resonance measurements are conducted in dark and under excitation similar to that used in the PL experiments and shed light on the possible origin of traps in the different samples. A simple model built on this concept agrees well with the experimentally-observed decay curves.

I. Introduction

Defects in silicon carbide (SiC) have attracted attention during the past few years from point of view of possible applications in quantum technologies. The latter include single-photon emitters applicable in quantum information processing,[1-5] magnetic sensors based on the Si vacancy in 4H-SiC,[6-8] and quantum bits (qubits).[9] Owing to its maturity, the 4H-polytype of silicon carbide (4H-SiC) is the most studied polytype so far. The divacancy in 4H-SiC, which is the main subject of this work, consists of carbon ($V_C$) and silicon ($V_{Si}$) vacancies positioned on neighbouring lattice sites (denoted hereafter as VV for brevity). In 4H-SiC, there exist four inequivalent configurations of $V_{Si}$-$V_C$ due to the presence of two inequivalent lattice sites for each vacancy in the unit cell.[10] The two inequivalent sites are usually termed hexagonal (denoted *h*) and cubic (*k*) depending on whether the arrangement of the next-nearest neighbours mimics that in the hexagonal würtzite or in the cubic zinc blende structure. Using the order $V_{Si}V_C$ for the site positions of the vacancies, the four inequivalent configurations are denoted further as *hh*, *kk*, *hk* and *kh*. When different charge states of any defect must be specified, we will use superscripts explicitly denoting the charge state, e.g., $VV^0$ denotes the neutral charge state of the divacancy, while $V_{Si}^{2-}$, for instance, denotes the double-negatively charged state of $V_{Si}$, etc.

The attractive properties of the divacancy in the 4H-SiC for solid-state quantum bit (qubit) have been demonstrated recently.[9] In this work, the authors associate the four lines observed in photoluminescence (PL) previously known as PL1 – PL4 with divacancies in each of the four inequivalent configurations: *hh* (PL1), *kk* (PL2), *hk* (PL3), and *kh* (PL4). *Ab initio* calculations[11] and electron paramagnetic resonance (EPR) measurements[12,10] show that the neutral divacancy has an electron spin $S = 1$ and the ground and excited states are triplets with $A_2$ symmetry in $C_{3v}$ ($^3A_2$). The defect can be spin-polarized using optical pumping[11] with millisecond spin coherence times[4] and a high-fidelity infrared spin-photon interface has been demonstrated.[13]

In the present work we address the quenching of the PL from the divacancies when certain photon energies are used for excitation. The quenching phenomenon is not unique for the divacancy and has been observed for several other defects at resonant as well as non-resonant excitation. Resonant excitation in this context means that the PL is excited at the zero-phonon transition energy and registered by observing the phonon sideband accompanying the zero-phonon line at lower photon energies. The quenching of the divacancy has been mentioned, e.g., in Ref. 14 (see the Supplementary Information in this reference). It has been shown that application of a second laser (repump laser) with higher photon energy and much lower power completely recovers the PL intensity of the divacancy (and other defects exhibiting quenching).[14] The quenching effect is also the most probable reason for the observation of the so-called blinking behaviour exhibited by single photon sources.[1] Although systematic investigation of the quenching effect has been missing until recently, researchers who have observed quenching tend to associate it with change of the charge state of the investigated defect.[14]

At the time of writing of this manuscript we have found two other papers[15,16] treating the quenching of the divacancy in 4H SiC. The main difference between these works and ours from experimental point of view is the use of microscope-based setup in Refs. [15,16], while in our work we use macroscopic lenses for focusing the lasers and collecting the signal. This leads to observation of much slower quenching dynamics in our case than when microscope objective is used. In our work, we combine photoluminescence time-decay measurements with photoluminescence excitation spectroscopy (PLE) and EPR measurements conducted at similar illumination conditions as the PL. This approach makes it possible to observe signatures in the EPR spectrum which change upon the infrared (IR) or repumping excitation on a similar time scale as observed in the PL. Using PLE we are also able to distinguish the threshold laser energies for "switching" on and off the quenching effect, which are found to differ for the divacancies with different inequivalent configurations. The experimentally-obtained thresholds are considered in the light of a new accurate theoretical calculation from first principles for the energy positions of the charge-transfer states of the divacancy. The improved accuracy of this calculation allows direct comparison of the results for the individual VV configurations with the experimental data. The good agreement between theory and experiment suggests that the observed quenching of the VV PL is due to conversion of the neutral divacancies to their negative charge state. Finally, we propose a dynamical model of the quenching which is simpler than the one reported in Ref. [15] and shows good agreement with experiment. Whenever appropriate, we compare our results and their interpretation with the results and discussion in the above-mentioned Refs. [15,16].

In Sec. II we describe the samples and the experimental details. For the rest of the scope we choose a style of presentation in which the experimental data is presented first (Section III) in order to set up foundation for the following discussion. In Sec. IV we present the new results of first-principles calculations on the charge transition levels of the divacancy, which are discussed further in Sec. V in conjunction with the experimental data in order to elucidate the underlying physics of the quenching effect. Finally, in Sec. VI we describe a dynamical model built on the ground of the notions developed in Sec. V, and test its viability in describing the quenching phenomenon. Sec. VII summarizes the conclusions.

All measurements in this work are done on ensemble of divacancies.

II. Samples and Experimental Details

In this study we present results from three 4H-SiC samples, all exhibiting strong luminescence from the VV centre (the PL1 – PL4 lines). However, the quenching dynamics of the divacancy photoluminescence (VV PL) in these samples is quite different; in particular, one of the samples does not exhibit quenching at all when excited with the same photon energies for which the other two samples do exhibit quenching. These differences in the quenching behaviour will be explained within the physical model treated later in Secs. V and VI.

One of the samples has two counterparts cut from the same high-purity semi-insulating (HPSI) 4H-SiC wafer, but irradiated to different doses, $10^{17}$ and $10^{18}$ cm$^{-2}$, with electrons of energy 2 MeV. The samples are subsequently annealed at a temperature of 800 °C for half an hour in order to create the divacancies. Most results presented here are from the piece irradiated to a dose of $10^{17}$ cm$^{-2}$, referred to further as SI1, although the other piece referred to as SI1 ($10^{18}$ cm$^{-2}$) has also been measured and shows similar results. The EPR results are obtained from this latter sample.

Another sample is n-type (N-doped) bulk 4H-SiC, referred to further as n-SiC. The N-doping level is in the low $10^{17}$ cm$^{-3}$ range. The specimen presented in this work is irradiated to $2\times10^{18}$ cm$^{-2}$ at 800 °C. Both SI1 and n-SiC samples exhibit strong quenching of the VV spectrum upon excitation with appropriate photon energy (e.g., 1.2 eV), but at quite different rates.

The third sample is an as-grown HPSI 4H-SiC substrate which, however, exhibits strong divacancy spectrum without any irradiation/annealing. This particular sample does not show any quenching in the PL lines from the divacancy at the excitation photon energies at which the VV-PL in the rest of the samples quenches. We refer to this sample further as SI2.

Most PL measurements are performed using a double monochromator (SPEX 1404) on the detection side. The monochromator is equipped with 600 grooves/mm gratings blazed at 1000 nm and an InGaAsP photomultiplier, which ensures optimum sensitivity in the emission region of the divacancy (wavelengths ~1078 – 1130 nm). The samples are mounted in a variable-temperature cryostat operated with liquid helium. We use a tuneable Ti-sapphire laser as an excitation source (referred to further as IR laser, or IR excitation), which can be combined with a coincident green laser (532 or 514.5 nm). The repump green laser (~ 10 mW power) is not focused on the sample and we estimate its power density to ~ 0.1 W/cm$^2$. Even at several orders of magnitude lower power than the IR excitation, the green laser has strong effect on the intensity of the divacancy emission and the quenching properties, although the green excitation alone is very inefficient in exciting the divacancy spectrum.

Using an IR-laser power of ~30 mW and moderate focusing to ~ 1 mm$^2$ spot on the sample, one estimates power density about 3 W/cm$^2$ (photon flux density ~ $1\cdot10^{19}$ cm$^{-2}$s$^{-1}$ at photon energy $h\nu$ = 1.2 eV). This is the main difference between our experimental conditions and these in Refs. [15,16], where the exciting laser has been focused by a microscope objective to a spot of the order of 1 μm$^2$. Thus, even if the laser power used with an objective is less than a milliwatt, the power density at the sample can be at least four orders of magnitude higher than ours due to the smaller spot. We will see later that this high-power excitation influences strongly the quenching dynamics by making it much faster than that observed in our work.

III. Experimental results

In this section, we present our experimental results, which will be discussed later in the light of the theoretical calculations presented in the Sec. IV. The following discussion in Sec. V describes the concepts used for interpretation of the experimental data and uses the theoretical results of Sec. IV to develop physical understanding of the quenching effect, which is used in Sec. VI to build a model for the PL decay. It will be shown that the experimental data together with the theoretical calculation strongly favour the model identifying the negatively-charged state of the divacancy as the "dark" state into which the divacancy is converted during quenching. This result agrees with [15], but is in contrast with the model presented in [16] suggesting the positively charged state as the "dark" state instead.

A. Optical properties

The general appearance of the PL spectra of the samples excited with photon energies 1.528 eV (811 nm) and 1.333 eV (930 nm) is shown in Fig. 1(a). Both these excitations are above the energy threshold for quenching for all divacancy configurations, but the former is chosen also above the threshold for excitation of the negatively-charged Si vacancy $V_{Si}^-$ (the zero phonon lines are V1 at 861.6 nm and V2 at 916.5 nm)[17] in order to examine its appearance in the samples.

All samples exhibit strong divacancy emission (the PL1 – PL4 lines), however, the appearance of the rest of the lines differs substantially between the spectra. Thus, the $V_{Si}^-$ spectrum (V1, V2 and associated phonon sidebands) appears only in the samples exhibiting quenching, SI1 and n-SiC, but is missing in the SI2 sample. The latter sample (SI2) shows weak contribution of a doublet around 1280 nm denoted as V in Fig. 1(b). This doublet is identified as the PL signature of vanadium in 4H-SiC,[18] and its presence may have impact on the lack of quenching of the VV PL in this sample, as discussed later in Sec. V. We notice also the appearance of the PL5 and PL6 lines observed previously in work investigating the divacancy properties.[9] These lines are well visible in the SI1 and SI2 samples, but not present at all in the n-SiC sample. Two more lines denoted here as PL5´ and PL6´ with intensity similar to that of PL5 and PL6 appear in all the three samples, but at different excitation conditions, as seen in Fig. 1. The origin of PL5, PL6, PL5´, and PL6´ is not known and these lines will not be discussed further.

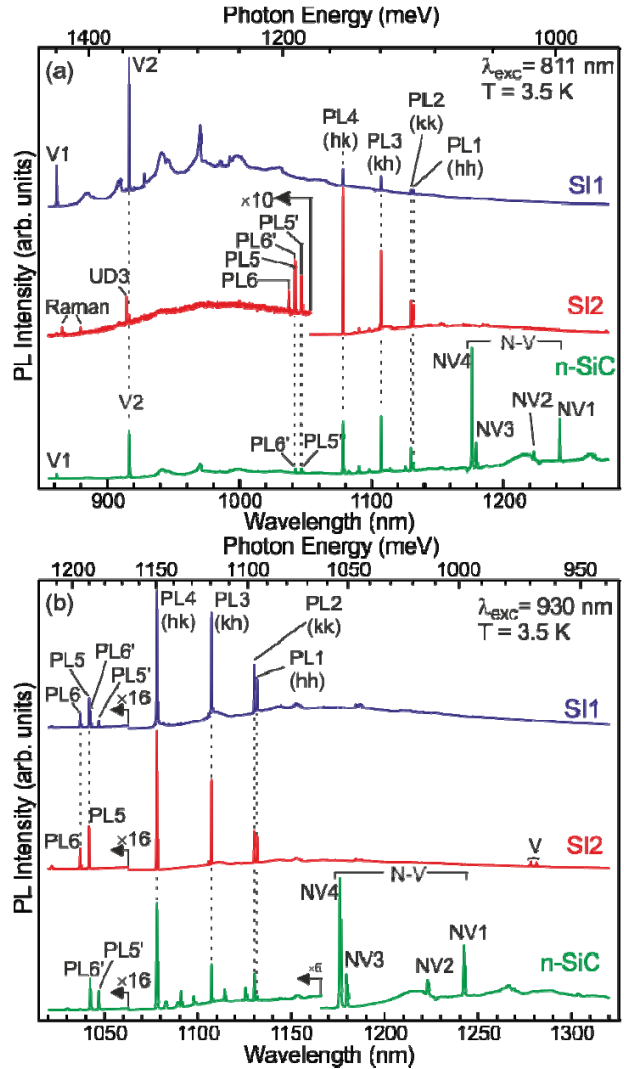

Fig. 1. PL spectra of the three investigated samples recorded with two different laser excitations, 811 nm (1.528 eV) and 930 nm (1.333 eV). Both excitations do not cause quenching of the VV-related lines (PL1 – PL4), but excitation with 811 nm allows also observation of the silicon-vacancy ($V_{Si}^-$) spectrum (the V1 and V2 lines and associated phonon sidebands), whenever present. 'Raman' denotes Raman lines. UD3 denotes an unidentified defect[19] which according to our recent unpublished data is most likely related to tantalum.[20] The SI2 sample exhibits weak contribution from vanadium (V), whereas the n-SiC sample shows strong contribution from a defect tentatively associated in recent work[23] with the nitrogen – vacancy pair in 4H-SiC.

Finally, we make a detour to comment on the four lines in the range ~ 1175 – 1245 nm denoted N–V in Fig. 1, which are observed only in the n-SiC sample. In an early work,[21] these lines were tentatively associated with the carbon vacancy – carbon antisite pair (CAV), in analogy with a work on 6H-SiC.[22] This latter work assigns six so-called P6/P7 lines observed in 6H-SiC in the range ~ 0.99 – 1.07 eV to the CAV defect in its six crystallographic-inequivalent configurations. Four inequivalent configurations are expected in 4H-SiC, hence the association made in [21] for the four lines appearing in 4H-SiC in the same spectral region as in 6H-SiC. Recent work,[23] however, suggests that the lines are instead due to the N-V pair, a defect expected to appear in SiC and to have properties similar to those of the well-studied N-V pair defect in diamond.[24] Our spectrum is identical to that of Ref. 21, and since these lines are observed only in the highly N-doped sample, we tend to confirm the association with the N-V centre made in Ref. 23. However, the interpretation of the spectrum observed in [23] does not seem to be entirely correct, and the following corrections are needed, in our opinion. First of all, the mutual disposition of the lines observed in [23] and ours is the same, apart from a common 0.6 meV shift for all lines, which can be attributed to slight miscalibration of their measurement and/or ours. Thus, there is no doubt that the same spectrum denoted N-V in Fig. 1 is observed also in Ref. 23. However, the low-energy peak denoted PLX1/PLX2 in Ref. 23 is claimed to comprise two lines, which are not resolved in their spectrum due to low spectral resolution (estimated to ~2 meV from their figures). In our measurement the low energy peak NV1 is a single symmetric line with apparent linewidth of 0.5 meV (our resolution is ~ 0.3 meV). The origin of the structure visible in the PLX1/PLX2 peak in Ref. 23 is not known, but such structure may arise, for instance, as a consequence of deteriorated focusing on different parts of the array detector used in this work. Furthermore, the high-energy peak denoted NV4 in Fig. 1 is observed also in Ref. 23, but incorrectly attributed to tungsten (W). In fact, their spectrum does display tungsten contribution, but the two W peaks separated by 1 meV according to the original work[25] are not resolved and constitute their highest-energy peak at ~ 1059 meV, whereas the line at ~ 1054 meV seen in their spectra actually belongs to the N–V spectrum, not to W. With these corrections in mind, our N–V spectrum is identical to that of [23]. We notice that with the above corrections the accuracy of agreement between the N-V lines and the theoretical results presented in [23] remains very good, within 50 meV for all the lines. A more recent calculation[26] provides even better agreement with our assignment for the NV1 – NV4 lines. Since the N-doping level in our sample (low $10^{17}$ cm$^{-3}$) is similar to that of the sample used in Ref. [23] ($2 \cdot 10^{17}$ cm$^{-3}$), and since these lines are not observed at all in the two HPSI samples, our result can be seen as a confirmation of the association of the considered lines with the N-V centre in 4H-SiC.

The PL peaks observed in the three samples are summarized in Table I.

Table I. List of the lines observed in PL in the three samples, in order of ascending wavelengths (descending energies). The association of UD3 with tantalum is tentative.[20] In the case of vanadium, only the stronger low-temperature components $\alpha_1$ and $\alpha_3$ are given. We follow the notations used for the components of the vanadium related $\alpha$-line in 6H-SiC.[27] The $\alpha_2$ and $\alpha_4$ components are also weakly visible at 3.5 K (blue-shifted by 0.8 meV from $\alpha_1$ and $\alpha_3$, respectively), but are omitted in the table.

| Line | Peak position in nm (meV) | Sample |
| --- | --- | --- |
| V1 ($V_{Si}$)[a] | 861.6 (1438.6) | SI1, n-SiC |
| UD3 (probably Ta)[b] | 914.5 (1355.4) | SI2 |
| V2 ($V_{Si}$)[a] | 916.5 (1352.4) | SI1, n-SiC |
| PL6[c] | 1037.7 (1194.5) | SI1, SI2 |
| PL5[c] | 1041.9 (1189.6) | SI1, SI2 |
| PL6´ | 1042.6 (1188.9) | SI1, n-SiC |
| PL5´ | 1047.3 (1183.5) | SI1, n-SiC |
| PL4 ($VV_{hk}$)[c] | 1078.5 (1149.3) | All samples |
| PL3 ($VV_{kh}$)[c] | 1107.6 (1119.1) | All samples |
| PL2 ($VV_{kk}$)[c] | 1130.5 (1096.4) | All samples |
| PL1 ($VV_{hh}$)[c] | 1132.0 (1095.0) | All samples |
| NV4[d] | 1176.4 (1053.6) | n-SiC |
| NV3[d] | 1180.0 (1050.4) | n-SiC |
| NV2[d] | 1223.2 (1013.3) | n-SiC |
| NV1[d] | 1242.8 (997.3) | n-SiC |
| Vanadium ($\alpha_3$ line)[e] | 1278.6 (969.4) | SI2 |
| Vanadium ($\alpha_1$ line)[e] | 1281.5 (967.2) | SI2 |

[a] From [17]
[b] See [20]
[c] From [9]
[d] New assignment based on our results and [23]
[e] Labelled in analogy with the $\alpha$-line in 6H-SiC [27]

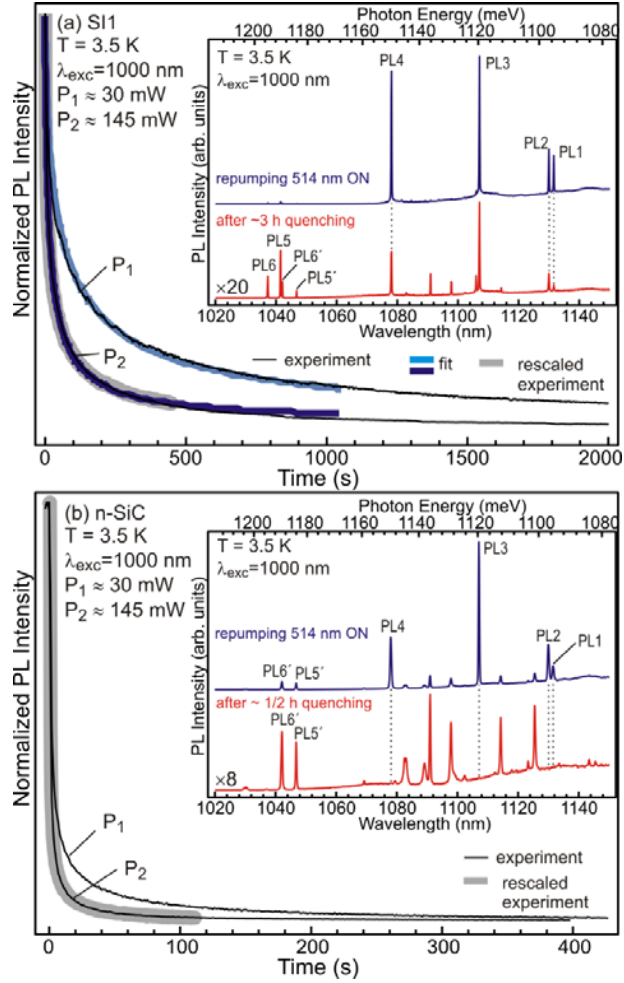

Fig. 2. Normalized decay curves illustrating the quenching of the PL4 line upon 1000 nm excitation at two different laser powers in (a) the SI1 and (b) the n-SiC sample. The insets show the corresponding spectra with repump laser at 514.5 nm (top curves) and after quenching with 1000 nm excitation for the indicated time. The intensities of the rest of the lines present in the spectra are not affected by the repump laser. Notice the different time scales and the different scaling factors for the bottom curves in the insets. The thick grey curves underlying the fast decay at short elapsed times in both panels (rescaled experiment) are obtained by scaling uniformly along the time axis the corresponding slow-decay curves corresponding to lower-power excitation, thus illustrating the scaling property discussed in text. The bold curves (magenta and blue in the online version) in panel (a) are fits obtained from the model presented in Sec. VI.

B. Quenching and recovery of the PL1 – PL4 lines

Fig. 2 displays a summary of the quenching behaviour of the divacancy lines observed in the SI1 and n-SiC samples with laser excitation at 1000 nm (1.24 eV). The SI2 sample does not show any quenching with this excitation. The PL spectra of SI1 and n-SiC displayed in the insets of panels (a) and (b) of the figure are obtained with and without repump laser at 514 nm (2.41 eV). When the repump laser is present no quenching is observed in any of the PL1 –

PL4 lines. However, if the repump laser is switched off, the quenching starts immediately, as illustrated by the decay curves in Fig. 2 (zero time corresponds to the switch-off moment). The figure illustrates also that the quenching rates are very different for the two samples SI1 and n-SiC (notice also the different time scales in the two panels). Thus, the spectrum for the SI1 sample in Fig. 2(b) is taken after ~ 3 hours of quenching, but it still shows contribution from the divacancy spectrum. On the other hand, all divacancy-related emission of the n-SiC sample quenches completely within about 10 minutes of irradiation with the 1.24 eV laser, as seen from the spectrum and shown for the PL4 line in Fig. 2(b).

In comparing our results with those presented in Refs. [15,16] we notice that the time scale for quenching of the VV PL is very different from ours. In both references the quenching requires sub-second times (milliseconds to tens of microseconds) for reaching a steady state (usually, close to zero PL intensity), whereas typical quenching times under the excitation conditions used in our work are tens of seconds and up to hours, depending on the sample. As already mentioned in Sec. II, the main difference between our measurements and these made in Refs. [15,16] is the laser power density on the sample. Therefore, we have investigated the power dependence of the quenching decay. The two decay curves in each panel of Fig.2(a,b) represent the normalized decay of the PL4 line with time obtained at two different exciting-laser power levels, $P_1 \approx 30$ and $P_2 \approx 145$ mW, thus $P_2/P_1 \approx 4.8$. Apparently, higher excitation power speeds up the quenching. However, we observe an interesting property concerning the power dependence of the decay. Namely, if the curve obtained at lower power is rescaled along the time axis by a certain factor, it coincides almost exactly with the curve obtained at higher power. The thick grey curves in Fig. 2(a,b) marked 'rescaled experiment' are the slow-decay curves rescaled by a factor ~1/4.55 (for Fig. 2(a)) and ~1/3.85 (for Fig. 2(b)); the rescaled curves match the corresponding fast-decay curves almost exactly. Since both rescaling factors are close to 1/4.8, which is the ratio between the low and high powers used, this experiment suggests that the quenching rate is roughly proportional to the exciting power. We notice that our simple model of the quenching dynamics discussed in more details later in Sec. VI reproduces quite accurately the just-described "scaling" property, as well as the shape of the decay for the SI1 sample. This is illustrated in Fig. 2(a) where the bold lines overlapping the decay curves in the range 0 – 1050 s represent the fits obtained from the model. We notice, however, that a good fit for the n-SiC sample (Fig. 2(b)) could not be obtained with the simple model considering only one type of traps, as discussed later in Sec. VI (see also the Supplemental Material (SM file)).[28]

Further properties of the quenching effect showing that in darkness the quenching state (partially or fully quenched PL level), as well as the recovering action of the repump laser are preserved ("remembered") for a long time at low temperatures, are described in more detail in the SM file (memory effects I and II).[28]

From the general behaviour of the VV PL in the three investigated samples it can be concluded that the observed quenching (or its lack) is associated with interaction with other defects, which are referred to as "traps" in the description of the dynamical model of the quenching presented later on. During this interaction effective under excitation the divacancy accumulates in a different charge state (positive or negative) and the luminescence from the neutral charge state (PL1 – PL4 lines) decays. Further, we will provide arguments that actually the negative charge state is the one that accumulates during quenching, which is in agreement with [15] and disagrees with [16].

C. Temperature dependence

The quenching-recovery properties of the VV PL in the samples exhibiting quenching are preserved also at higher temperatures. Let us consider first the temperature dependence of the PL spectra displayed in Fig. 3. Apparently, the PL3 line dominates the spectrum at temperatures above ~ 60 K and above approximately 150 K all contribution from the sharp PL1 – PL4 lines becomes indistinguishable. It is also interesting to compare the higher-temperature spectra of the three samples. The spectra of the three samples excited with 930 nm at 200 K are compared in the upper inset of Fig. 3. It is quite obvious that the only sample the high-temperature emission of which stems from the divacancy is the SI2 sample, which does not exhibit any quenching. In the samples exhibiting quenching, on the other hand, we see that the spectra at 200 K are dominated either by the silicon vacancy $V_{Si}^-$ (SI1), or by the N–V – associated band at lower energies (n-SiC), whereas the VV PL contribution is negligible. The lower inset in Fig. 3 illustrates that the VV emission in the SI1 sample can become dominant at 200 K when lower-energy excitation is used (1000 nm), but this contribution still vanishes at room temperature (293 K).

We notice also the obvious up-conversion expressed in the fact that the $V_{Si}^-$ spectrum dominates the emission of the SI1 sample despite of the fact that the IR excitation energy in both insets of Fig. 3 is *below* the energy of both zero phonon lines of $V_{Si}^-$, V1 and V2. The up-conversion becomes apparent at temperatures above ~80 K. The up-conversion mechanism is discussed later in Sec. V, where we show that the observed up-conversion also supports the identification of the negative charge state of the divacancy as the dark state. However, it is quite clear that in both samples exhibiting quenching the VV spectrum nearly vanishes above ~ 150 K and the remaining dominant emission is due to other defects ($V_{Si}^-$ in sample SI1, or the N–V pair in sample n-SiC). This observation may explain why the effect of repumping diminishes and vanishes at higher temperatures above ~ 150 K with the excitation of 976 nm (1.27 eV) reported in [15]. This effect may simply be due to vanishing contribution of VV in the spectrum, but no spectral data is given for elevated temperatures in Ref. 15. We will discuss more on this subject in Sec. V.

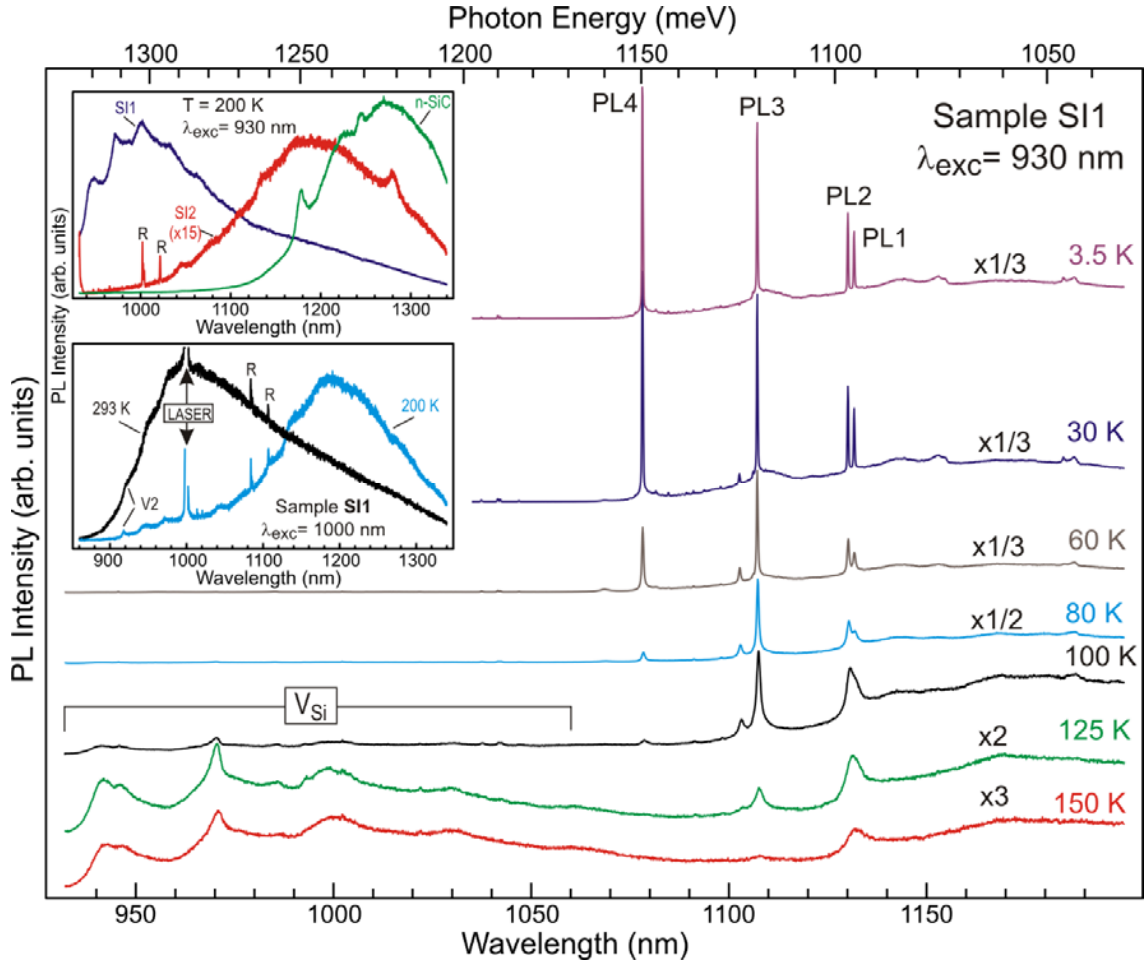

Fig. 3. Temperature dependence of the VV spectrum in the SI1 sample with 930 nm excitation. The upper inset compares the spectra of the three samples at 200 K, illustrating that VV contribution at this temperature and with 930 nm excitation is significant only in the SI2 sample which shows no quenching. The lower inset compares the spectra obtained with 1000 nm excitation of the SI1 sample at 200 K and room temperature. This inset illustrates the upconversion in excitation of $V_{Si}$, and also that the VV spectrum can be emphasized with 1000 nm excitation at 200 K, but still vanishes at room temperature. 'R' denotes Raman lines. Note the scale changes applied to some spectra in order to make them commensurate for display.

Further details on the quenching-recovery behaviour of the VV PL at different temperatures are given in the SM file.[28]

D. EPR results

We turn now to summarizing the EPR results. In order to reproduce closely the conditions of the PL measurements we have recorded EPR spectra in darkness, or under illumination with a 1030 nm (~1.2 eV) diode laser. A 532 nm diode laser or a white LED is used as a replacement for the repump laser. The observed EPR spectra measured in darkness and under illumination in samples SI1 ($10^{18}$ cm$^{-2}$), n-SiC, and SI2 are shown in Fig. 4. We notice that the EPR spectrum of $V_{Si}^-$ also appears in the former two samples, but under different phase conditions (not shown). The SI2 sample does not display contribution from $V_{Si}^-$, in agreement with the PL spectra in Fig. 1.

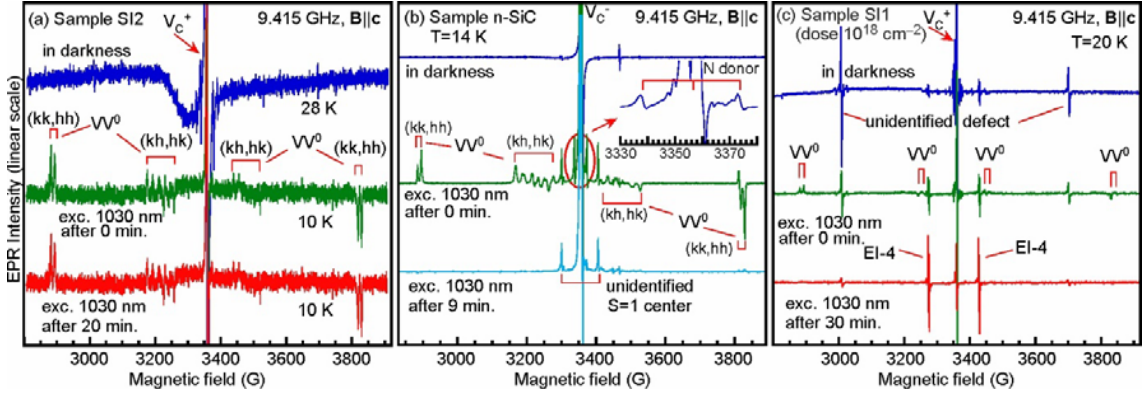

Fig. 4. EPR spectra of the samples in darkness and after illumination with 1030-nm laser for various durations, as indicated for each curve. The spectra in part (a) are from the SI2 sample, in part (b) from the n-SiC sample, and in part (c) from the electron-irradiated HPSI 4H-SiC (same substrate as SI1, but irradiation dose $1 \cdot 10^{18}$ cm$^{-2}$). The middle lines for the spectra in darkness in panels (a) and (c) are from the positively-charged C vacancy ($V_C^+$),[29,30] and in panel (b) from the negatively charged one ($V_C^-$).[31-33] The inset in part (b) illustrates the appearance of the N shallow donor spectrum in the beginning of the IR excitation; this spectrum vanishes in synchrony with the neutral divacancy quenching. The signals of axial (kk and hh) and basal (kh, hk) configurations of the neutral divacancy ($VV^0$), as well as EI-4[33] and another unidentified defect with spin one are indicated.

Let us consider first the SI2 sample (no VV PL quenching). The spectrum in dark (Fig. 4(a)) shows only contribution from the positively-charged carbon vacancy $V_C^+$,[29,30] but illumination with either 1030 nm or white LED (or both) brings up the divacancy spectrum (the lines denoted $VV^0$ in Fig. 4). The appearance of the divacancy lines in EPR only under illumination may be due to two different mechanisms. The first one is that merely the divacancies are in their positive charge state and are driven to their neutral charge state by 1.2 eV illumination (cf. the energy diagram in Fig. 6). The second mechanism involves the spin-polarization properties of the divacancy,[11] as explained in the following. The ground state of the divacancy is a triplet $^3A_2$, which splits into $M_S = 0$ and $M_S = \pm 1$ substates under magnetic field ($M_S$ denotes the magnetic quantum number). Optical excitation and subsequent recombination leads to preferable population of the $M_S = 0$ spin state.[11] This increases the EPR signal which depends on the relative population of the $M_S = 0$ and $M_S = \pm 1$ states. Both mechanisms can be effective for activating the EPR signal of VV. However, an equilibrium concentration of neutral divacancies is almost instantly established after illumination and, as we will discuss later in Sec. V, there are no reasons for decay of this equilibrium concentration.

On the other hand, the EPR spectrum of the n-SiC sample in darkness shows dominating contribution from the negatively-charged carbon vacancy ($V_C^-$).[31-33] The $VV^0$ signal appears after illumination with 1030-nm laser, but decreases with time and vanishes within ~ 10 min, as illustrated by the sequentially-recorded spectra in Fig. 4(b). We notice also the appearance initially upon 1030-nm excitation of the EPR signal from the neutral shallow N-donor (inset in Fig. 4(b)). This signal also vanishes within ~10 min from the beginning of the illumination.

The appearance of the neutral N-donor EPR spectrum upon illumination can be seen as an indicator for the presence of free electrons in the conduction band (CB), as will be discussed further in Sec. V.

Finally, we consider the EPR spectra of the SI1 sample electron-irradiated to $10^{17}$ cm$^{-2}$ and its counterpart irradiated to $10^{18}$ cm$^{-2}$. [In the following discussion we will distinguish them as SI1 ($10^{17}$ cm$^{-2}$) and SI1 ($10^{18}$ cm$^{-2}$)]. Similar to the other two samples, the observation of the VV$^0$ spectrum in EPR requires application of at least 1030 nm excitation. However, while the PL quenching properties of these two samples are very similar, we have succeeded in observing quenching of the VV$^0$ EPR spectrum only in the sample irradiated to $10^{18}$ cm$^{-2}$. In this case, illustrated in Fig. 4(c), the quenching occurs on similar time scale as that for the luminescence, indicating that the divacancy ensemble is eventually transferred into different charge state. However, in the SI1 sample irradiated to $10^{17}$ cm$^{-2}$ we could not capture the quenching of the EPR signal from VV$^0$ under excitation with 1030 nm (not shown). The reasons for this apparent discrepancy with the PL behaviour are not understood at present. However, there are two arguments favouring the possibility that the VV$^0$ concentration reaches equilibrium low level already before the first EPR scan is taken. Indeed, the quenching of the VV$^0$-related lines in PL of this sample is never complete even after hours of illumination with 1.2 eV photons, indicating that the equilibrium concentration of VV$^0$ after long-term quenching is nonzero (in contrast, e.g., to the n-SiC sample, where the quenching is complete). The second argument is based on the observation that the VV$^0$ spectrum increases strongly when repumping excitation is applied together with the IR excitation. In fact, the VV$^0$ EPR signal increases also in the n-SiC and in the SI1 irradiated to $10^{18}$ cm$^{-2}$ samples when the repumping excitation is applied, but in these cases the observed intensity of the VV$^0$ EPR lines is close to that observed in the beginning of the quenching. In contrast, in the SI1 ($10^{17}$ cm$^{-2}$) the observed enhancement is more than five times the intensity observed initially upon 1030 nm excitation. Both these arguments support the view that the equilibrium low concentration of the VV$^0$ is observed in the SI1 ($10^{17}$ cm$^{-2}$) sample, but the reasons for the apparent discrepancy with the time scale of the PL quenching remain unclear.

The EPR spectrum of the SI1 ($10^{18}$ cm$^{-2}$) sample in dark displays clear contributions from the positively-charged carbon vacancy $V_C^+$, the EI-4 centre,[34] and another unidentified spin-one defect, as indicated in Fig. 4(c). Upon 1030 nm illumination the VV$^0$ spectrum appears, but quenches below the EPR detection limit in about 30 min. The intensities of EI-4 and the unknown spin-one defect also change upon prolonged 1030-nm excitation, as seen in Fig. 4(c).

E. Photoluminescence excitation spectroscopy

We have performed photoluminescence excitation (PLE) spectroscopy on the two samples exhibiting quenching (SI1 and n-SiC) in order to determine the threshold excitation energies below which quenching is observed, but above which the PL intensity is stable. The threshold energies are different for the different lines (PL1 – PL4). In these experiments, the photoluminescence is monitored at the chosen line while the excitation-laser energy is scanned from low to high energies in the range approximately 1240 – 1385 meV (1000 – 896 nm). The results are displayed in Fig. 5.

A remark is due concerning the existence of what appears to be excessive noise in the PLE spectra in Fig. 5. We have taken care to maintain approximately constant power of the exciting IR laser during the automated scanning, but apparently most of the spikes observed in the curves can still be attributed to variations in the laser power. There is yet another reason for fluctuations in the PL signal, namely, variations in the laser pointing direction when the laser wavelength is scanned. The reason for minute laser-beam position variations on the sample is the spatial filter-monochromator which filters out the superluminescence from the exciting Ti-sapphire laser, and which is scanned automatically together with the laser. Thus, small variations in the excited volume of the sample may occur and since different volumes may have slightly different luminescent properties and different level of quenching, hence, variations in the PLE signal may be due also to variations in the laser beam position.

Despite these artificial spikes, clear thresholds can be observed for each of the four zero phonon lines of VV (PL1 – PL4) in the PLE spectra of the SI1 sample, see Fig. 5(a). The thresholds are marked with vertical lines and labelled with the corresponding energy of the threshold in eV (Fig.5). Laser energies below these thresholds cause quenching of the corresponding line, but above the threshold the intensity of the corresponding line is stable with time. It should be noticed that the lowest-energy threshold (of the PL4 line) appears also in the PLE spectra of the other lines (PL1 – PL3), because the latter three lines reside on the PL4-phonon sideband, which has quenching properties similar to that of the zero phonon line.

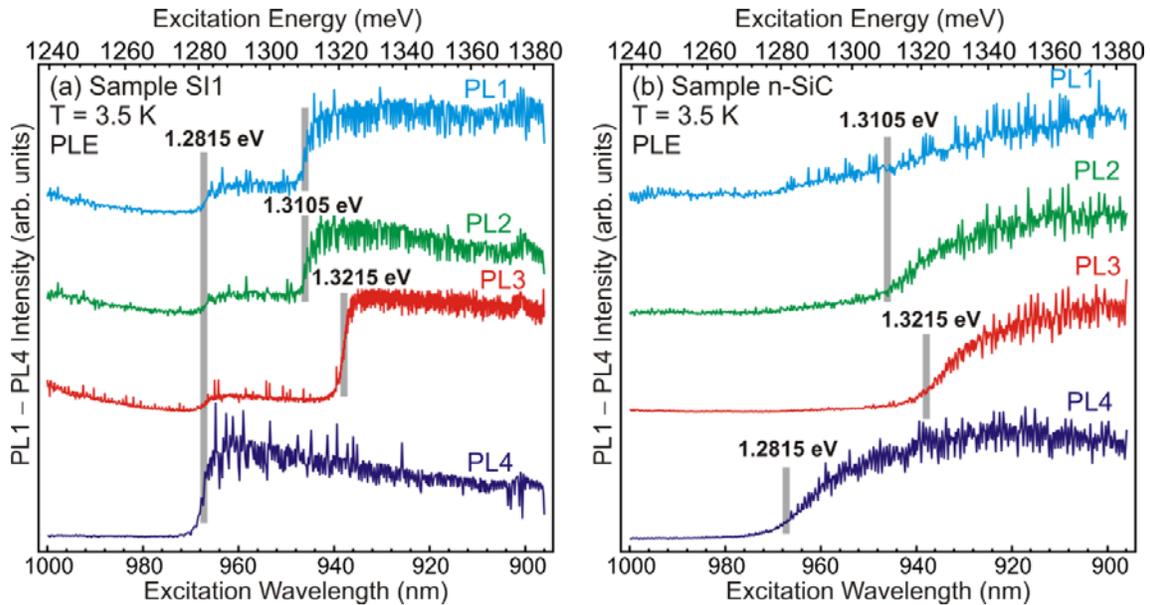

Fig. 5. PLE spectra at T = 3.5 K of the two samples showing quenching of the $VV^0$ luminescence: (a) the SI1, and (b) the n-SiC sample. The four curves in each panel correspond to monitoring each of the four zero phonon lines associated with $VV^0$ (PL1 – PL4), as denoted for each curve. The threshold energies which are much more distinct in the high-purity SI1 sample are denoted also in the graphs for the n-SiC sample, where they occur at the low-energy end of the transition between quenching and non-quenching behaviour of each line.

The thresholds observed in Fig. 5(a) are marked also in Fig. 5(b). While the n-SiC sample exhibits essentially the same threshold energies as the SI1 sample, the transitions between quenching – not quenching in this sample seem to be significantly broader. The observed broadening is most likely associated with inhomogeneous broadening of the charge-transition levels of the divacancies resulting from inhomogeneous distribution of the defects in the samples. In other words, the local environments of the individual divacancies differ leading to small shifts in their energy levels. These variations in the local environment are expected to be larger in the highly-irradiated and relatively high-doped n-SiC sample, resulting in larger broadening ($\geq$ 10 meV) of the transition between quenching and stable PL, compared to the sharper thresholds of ~ 4 meV in SI1, as estimated from Fig. 5(a). The notion of inhomogeneous broadening is corroborated by the observation of the PL temporal behaviour in the n-SiC sample when the exciting laser is tuned within the threshold-energy interval. Such a measurement (not shown here) shows stable non-vanishing PL intensity, but at a level lower than the maximum obtained with a few meV larger photon energy (recall that the VV PL vanishes in a matter of minutes for excitation below the threshold for this sample). This observation suggests that a part of the neutral-divacancy population is maintained stable because this part is capable of re-exciting captured electrons to the CB via absorption of photons, while the rest of the neutral divacancies convert to a dark state under this particular illumination and their luminescence is quenched. Thus, the observed broadening of the transition between quenching and non-quenching is attributed to inhomogeneous broadening of the charge-transition levels. We notice that our results agree with the PLE data presented in Ref. 16 (cf. their figure 2), albeit the exact threshold energies are not listed in this reference, and their conclusion that the dark state is the positively-charged state of the divacancy is opposite to ours.

Finally, we mention for completeness that the PLE spectra on the VV$^0$ lines of the sample without quenching (SI2) do not display any threshold-like features, as expected. We notice further that if the PLE spectra of the samples which do exhibit quenching (e.g., SI1) are measured when the repump laser is applied, then the obtained PLE spectra also do not exhibit thresholds and are similar to those of the "non-quenching" sample SI2 (see the SM file).[28]

IV. Results from first-principles calculations

Previously reported results[35-37] from first-principles calculations on the charge-transfer levels of the divacancy suggest that this defect has three charge states within the band gap of 4H-SiC: neutral (+/0), single-negatively charged (0/−), and double-negatively charged (−/2−), valid for all four inequivalent configurations of VV (*hk*, *kh*, *hh*, and *kk*). According to the calculations presented here, as well as previously reported data,[37] the energy separation between the (+/0) charge transition level and the VB edge (~ 1.1 eV) is slightly smaller than that between the (0/−) level and the CB edge (~ 1.2 – 1.3 eV), for all divacancy configurations. Thus, there exists a range of photon energies for which the positively-charged state can be converted into neutral by photoionization, but the negatively-charged state cannot be ionized by emission of electrons to the CB via absorption of the same photons. This notion is depicted also qualitatively in Fig. 6, which presents an approximate energy diagram for the charge-transfer levels of the divacancy (and other relevant defects), but without regard for the different inequivalent configurations. Consequently, one may assume that the negatively-charged state is the dark state into which the divacancies are converted during quenching (e.g., with 1.2 eV photons). In order to confirm this assumption we have conducted converging *ab initio* calculations for all four inequivalent divacancy configurations of the negatively- and positively-charged states. The improved accuracy of the present calculation is of crucial importance for the data interpretation, and the theoretical results can be compared directly with the threshold energies found from the PLE measurements for the different divacancy configurations.

The photoionization threshold of VV⁻ in 4H SiC can be estimated by the (0/-) charge transition level with respect to the conduction band edge (CBM). The adiabatic charge transition levels ($E_{q+1/q}$) can be calculated according to the following equation

$$E_{q+1/q} = E_{tot}^q - E_{tot}^{q+1} + \Delta V(q) - \Delta V(q+1), \tag{1}$$

where $E_{tot}^q$ and $E_{tot}^{q+1}$ stand for the total energy of the system in the $q$ and $q+1$ charge states, respectively, $\Delta V(q)$ and $\Delta V(q+1)$ are the charge correction terms of those. Thus, we calculate this level for all the possible defect configurations, and we also provide results for the (+/0) level. Calculations were carried out by means of HSE06 range-separated hybrid functional developed by Heyd, Scuseria and Ernzerhof.[38] In order to reach sufficient accuracy, we employed 2 x 2 x 2 Monkhorst-Pack k-point mesh[39] for this calculation in a 576-atom supercell. We notice in passing that if only G-point sampling in the k-space is used, as often encountered in other calculations on divacancies, the accuracy is not sufficient to distinguish the correct order of the energies of the different divacancy configurations. Thus, employment of a mesh in the k-space appears to be crucial for obtaining results accurate enough to be compared with the experimental data. Plane wave basis set with a cutoff of 420 eV was employed. Fully relaxed geometries were achieved by setting the force threshold of 0.01 eV/Å. Appropriate treatment of core electrons was provided by applying of projector-augmented wave (PAW) method.[40] Freysoldt charge correction[41] in total energy was used. The results are presented in Table II together with the experimentally-determined threshold energies.

TABLE II. Positions of (0/-) and (+/0) charge transition levels of $V_{Si}V_C$ defect configurations referenced to the CBM and VBM, respectively. In the calculations experimental value of the band gap of 3.285 eV at low temperature[42,43] was considered. Charge corrected values (Ē) and those neglecting the correction terms (E) are also presented. Experimental threshold energies of photoionization are provided.

| Line | Config. | Threshold (exp.) | $E_{0/-}^{CBM}$ (eV) | $\bar{E}_{0/-}^{CBM}$ (eV) | $E_{+/0}^{VBM}$ (eV) | $\bar{E}_{+/0}^{VBM}$ (eV) |
|------|---------|------------------|----------------------|----------------------------|----------------------|----------------------------|
| PL1  | hh      | 1.310            | 1.429                | 1.245                      | 1.131                | 1.070                      |
| PL2  | kk      | 1.310            | 1.386                | 1.209                      | 1.067                | 1.010                      |
| PL3  | hk      | 1.321            | 1.456                | 1.307                      | 1.108                | 1.051                      |
| PL4  | kh      | 1.281            | 1.345                | 1.174                      | 1.136                | 1.081                      |

One can see in Table II that the calculated (0/-) charge transition level with charge correction (Ē) are systematically lower by 0.02 – 0.11 eV than the experimental threshold energies. Nevertheless, the calculated photoionization thresholds are in the correct order, with the lowest threshold corresponding to PL4 and the highest – to PL3 and, in general, the theoretical values agree very well with the experimental ones. Furthermore, all values of $\bar{E}_{0/-}^{CBM}$ are higher than and well separated from those of $\bar{E}_{+/0}^{VBM}$, which are all below 1.1 eV in each configuration. This fact supports our qualitative explanation for the quenching effect presented in the next section, that is, the concentration of negatively charged divacancies is robust and conversion of neutral divacancy to negatively-charged cannot be reversed by excitation energy of ~ 1.28 eV (and below). On the other hand, positively charged divacancy can be converted to neutral by this excitation by promoting an electron from the VB to the degenerate *e* state lying in the band gap (or emitting a hole to the VB). Thus, the calculation results corroborate the concept that the experimentally observed thresholds are associated with photoionization of $VV^-$ to $VV^0$.

The important role of the employed charge correction method[41] is also revealed by the calculation results. While the charge corrected values ($\bar{E}_{0/-}^{CBM}$) are lower by up to 0.1 eV than the observed photoionization thresholds, the uncorrected values ($E_{0/-}^{CBM}$) are higher than that by about the same amount. Possibly, this indicates that the applied Freysoldt correction overestimates the correction to the total energy of the negatively charged defects in the 576-atom supercell.

V. Discussion

Let us discuss now qualitatively the observed quenching/recovery behaviour of the divacancy adopting the notion of charge transfer to/from other defects referred to in the following as 'traps'. In the course of the discussion, we should be able also to provide an explanation for the lack of quenching in the SI2 sample. We refer in the following to the energy diagram shown in Fig. 6, which contains a summary of the energy levels associated with the different charge states of $V_{Si}$, VV, $V_C$, and $V_CC_{Si}$. The energy positions for $V_{Si}$, $V_C$ and N–V are taken from literature,[26,33,44,45] whereas the results for VV are originally presented here (see Sec. IV). The energy levels designed in Fig. 6 are 'generic' average levels ignoring the site dependence, which is sufficient for a qualitative discussion. We notice that the photoionization of the shallow donors (N) and acceptors (mainly B) with 1.2 eV excitation is readily available. As already discussed, we adopt also the concept that the quenching of the divacancy-related PL with ~1.2 eV excitation is due to conversion of the divacancy to the negatively-charged state (0/−), because this concept agrees with the theoretical results from the previous section and the energy diagram in Fig. 6.

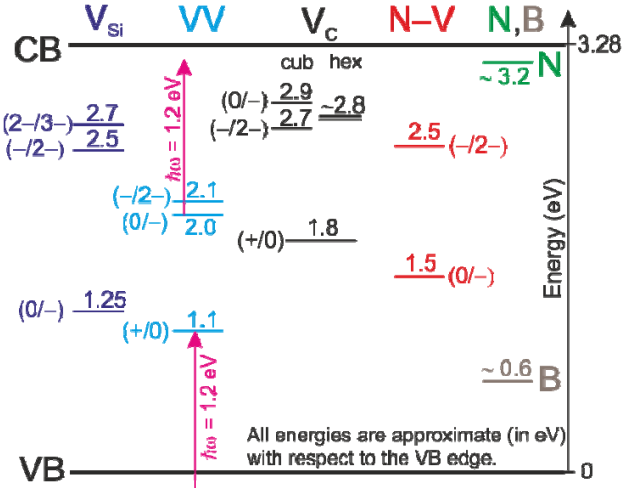

Fig. 6. Energy level diagram for the different charge states of $V_{Si}$, VV, $V_C$ and the N–V pair in 4H-SiC. Only approximate (generic) levels within the band gap are shown and labelled with approximate energy with respect to the valence band edge, without regard for the site dependence except for the single- and double-negatively charged states of $V_C$ where cubic (cub) and hexagonal (hex) sites are distinguished. The energy levels are drawn to scale within the band gap of 4H-SiC taken as 3.28 eV. N and B denote the nitrogen and boron shallow dopants, respectively. The two arrows illustrate action of photons with energy $\hbar\omega$ = 1.2 eV. VB and CB denote the valence and conduction bands, respectively.

In our further discussion we assume that co-existence of different charge states of the considered defects in different microscopic volumes of the sample is possible. This notion is based on the possibility for variations in the local environment of individual defects arising from variations in the local concentrations of different defects in microscopic volumes in the sample. In other words, we distinguish *average* in macroscopic volume defect concentrations which can be handled by means of the Fermi-level concept, from *local* concentrations, which may differ significantly from the average ones in microscopic volumes of the crystal. In addition to the deep-level defects, there are also shallow (unintentional) dopants, for instance N and B. Thus, it is known that the HPSI SiC contains shallow donors (N) and acceptors (B) in comparable concentrations, both in the $10^{15}$ cm$^{-3}$ range.[46] The unintentional shallow dopants in HPSI SiC partially mutually compensate each other, but any residual donors/acceptors are compensated by intrinsic defects, mainly $V_C$, $V_{Si}$ and VV. Thus, the shallow donors (N) and acceptors (B) in HPSI SiC are mostly ionized in equilibrium. The concentrations of the intrinsic defects observed in our samples and depicted in Fig. 6 are not known exactly, but we may assume that these concentrations are of the order of magnitude of those of the shallow-dopant concentrations ($\sim 10^{15}$ cm$^{-3}$).

When optical excitation of certain energy is applied to the crystal, some of the defects can be photoionized by electron (or hole) emission to the CB (or VB), respectively. The generated free carriers can then be recaptured to other defects. Thus, the populations of the various charge states change from their values in dark and reach a new equilibrium distribution upon continuous illumination. In discussing this new equilibrium distribution, we will distinguish qualitatively between *slow* and *fast* processes. We assume that emission of electrons to the CB or holes to the VB due to photoionization is a fast process, whereas the following re-capturing of the free carriers may be slow or fast depending on the charge state of the capturing defect. Thus, for instance, neutral defects are anticipated to have much smaller capture cross sections for the corresponding carriers than positively/negatively charged centres (for electrons/holes, respectively), owing to the giant capture cross section of charged defects associated with their long-range attractive Coulomb potential.[47,48]

We now consider the particular case of the divacancy defect under 1.2 eV excitation. Let us assume for the moment that there exists some population of positively-charged divacancies. Then the 1.2-eV excitation will rapidly convert $VV^+$ to $VV^0$ by promoting electrons from the VB to the (+/0) level (or, equivalently, by emitting holes to the VB). In principle, the free holes can be re-captured back to $VV^0$. However, in terms of our qualitative distinction between fast and slow processes, the conversion of $VV^+$ to $VV^0$ is fast, whereas the capture of holes by $VV^0$ to produce $VV^+$ again (the inverse process) is slow, because the capturing occurs to neutral divacancies. Let us assume now also that negatively-charged silicon vacancies, $V_{Si}^-$, are present (in accord with the PL spectra of SI1 and n-SiC in Fig. 1). It is then clear that the capture of the free holes is much more probable to occur to $V_{Si}^-$ (fast process) rather than to $VV^0$ (slow process). Other negatively-charged defects may also contribute to the (fast) capture of free holes, especially negatively-charged boron acceptors $B^-$, but also such as $V_C^-$ and $V_C^{2-}$. The latter processes are much more probable (much faster) owing to the long-range attractive Coulomb force, as discussed in the previous paragraph. Thus, the net effect of the 1.2 eV excitation on the divacancy is increase of the population of $VV^0$ at the expense of decreasing population of $VV^+$. We notice that photoionization of shallow uncompensated (neutral) B acceptors may also influence the total population of free holes, but in this case both the photoemission and the capture processes are fast according to our concept that negatively-charged $B^-$ impurities will have large capture cross section for holes. Within the described scenario, the concentration of the negatively-charged silicon vacancy $V_{Si}^-$ can only decrease at the expense of increasing $V_{Si}^0$ concentration.

In accord with our assumption that different charge states of the same defect are possible, we have to consider also the possible generation of free electrons in the CB due to photoionization with 1.2 eV optical excitation. Free electrons can be generated, e.g., by photoionization of the relatively shallow $V_C^-$ and $V_C^{2-}$. The higher negatively-charged states of $V_{Si}$ as well as the N–V$^-$ defect (if present) may also contribute to the generation of free electrons, as well as uncompensated (i.e., neutral) shallow N donors. However, as long as 1.2 eV excitation is considered, the photon energy is not sufficient to induce transitions from the (0/−) level corresponding to $VV^-$, hence the concentration of negatively-charged divacancies is not affected by the 1.2-eV illumination. In fact, the concentration of $VV^-$ can only increase on account of the *slow* capture of free electrons generated by other defects. The process is classified as slow, because the capture occurs to a neutral defect. We should mention also the possibility of $VV^-$ to capture holes, which in our scenario would be a fast process. However, in the considered scenario the only source of free holes is due mainly to conversion of positively-charged divacancies to neutral ones, and there are several possible traps for holes other than the negatively-charged vacancies ($V_{Si}^-$, $B^-$, $V_C^-$ and $V_C^{2-}$). Thus, quasi-equilibrium concentration of all defects considered in Fig. 6 can be established on a time scale shorter than that for accumulation of the negatively-charged divacancies, resulting in slow decay of the population of neutral divacancies and, consequently, the $VV^0$ luminescence. Thus, we associate the negative charge state with the dark state of the divacancy, in agreement with Ref. 15. The conversion of $VV^0$ to $VV^-$ is irreversible at low temperature, if only 1.2 eV excitation is applied.

We will now argue that the qualitative division of the light-induced processes into "slow" and "fast", as well as the notion of coexistence of different charge states of certain defect leading to the possibility of simultaneous generation of free electrons and holes, are actually in agreement with the experimental observations. Indeed, photoionization, i.e., emission of bound carriers to the corresponding bands is considered as a fast process, whereas capture to neutral defects is slow process. Thus, application of 1.2 eV illumination leads to slow quenching of the $VV^0$ PL owing to the slow conversion to $VV^-$, a consequence of the slow but irreversible capture of free electrons to $VV^0$. On the other hand, application of repump laser with photon energy > 1.3 eV causes photoionization of $VV^-$ and recovers the $VV^0$ population. This is a fast process, hence the ignition of repumping photons appears to recover the VV PL instantly in our measurements (see also the SM file).[28] We recall here that the PL recovery is not exactly instant, of course, but it is too fast to be captured with our limited time resolution (of the order of 1 second). However, the results presented in two other related studies[15,16] show that the rise time of the VV PL is finite and much faster than the time for quenching.

Furthermore, we notice that the observed upconversion of the silicon-vacancy PL ($V_{Si}^-$) can be seen as confirmation for the generation of both types of free carriers, electrons and holes. (Weaker upconversion, which is not shown, is observed also in the n-SiC sample.) Indeed, we assumed generation of free holes as a consequence of conversion of positively-charged divacancies to neutral ones. When the free holes are captured to the negative $V_{Si}^-$ neutral silicon vacancies $V_{Si}^0$ are generated, which can capture electrons. If the electron is captured to the excited state of $V_{Si}^-$ instead of to the ground state, it will return to the ground state radiatively and contribute to the $V_{Si}^-$ luminescence. Thus, we obtain upconversion but the electrons and holes necessary to maintain this upconverted luminescence are generated from different defects. Hence, the upconversion comprises fast capture of a hole followed by a slow capture of electron to the excited state of $V_{Si}^-$. The rate of the latter "slow" process depends on the free-electron concentration, so that the upconverted PL is proportional to the free electron concentration. Now we can comment on the role of the temperature in the upconversion. Indeed, the upconversion becomes more pronounced at elevated temperatures (cf. Fig. 3), and two possible reasons for that can be envisaged.

First, even if all the shallow N donors are ionized in dark (no free-carrier generation), the generation of free electrons due to photoionization of any other defects will establish a quasi-equilibrium neutral-donor concentration, because the shallow donors are the most probable traps for the free electrons owing to their giant capture cross section.[47,48]

The inverse process, photoionization of the neutral donors, also takes place and a quasi-equilibrium concentration of neutral donors and free electrons is rapidly established, which will be maintained by the continuous incident illumination. At elevated temperatures the thermal ionization of the donors also becomes significant. Thus, increased free-electron concentration is anticipated at higher temperatures, leading to increased upconverted PL from $V_{Si}^-$, as observed experimentally. The appearance of the N-donor spectrum in EPR in the n-SiC sample corroborates this idea (as already mentioned, the EPR signal from neutral donors can serve as a monitor of the concentration of free electrons in this sample). We notice also that the EPR signal from the neutral N donors quenches at a similar rate as the EPR signal from VV$^0$, thus indicating that, indeed, the free-electron concentration diminishes in synchrony with the conversion of VV$^0$ to VV$^-$. Similar effect is expected in the SI1 sample, but the donor concentration in this case is not sufficient to observe EPR signal.

The second possible reason for increasing of the V$_{Si}^-$ PL with increased temperature is the possibility for phonon-assisted photoionization of VV$^-$. We mean a process in which a photon and a phonon are absorbed to promote an electron from the (0/–) level of the VV (i.e., from VV$^-$) to the CB. Phonons must be available for absorption; therefore, the process will be significant only at elevated temperatures. This phonon-assisted absorption will also contribute to increased free-electron concentration thus promoting the upconverted PL of V$_{Si}^-$. In addition, the phonon-assisted photoionization process provides also a path for conversion of VV$^-$ to VV$^0$, thus explaining why the contrast between the "bright" and "dark" states of the divacancy PL decreases at elevated temperatures, as observed in [15].

We notice further that the energy conservation requires that, if phonons of energy $\hbar\Omega$ are involved in the phonon-assisted absorption process, then the condition

$$\hbar\omega \geq \Delta E - \hbar\Omega \tag{2}$$

must be fulfilled. Here $\hbar\omega$ denotes the laser excitation energy, and $\Delta E$ is the experimental energy separation between the CB edge and the (0/–) level of a given divacancy configuration (denoted as "Threshold (exp.)" in Table II). Considering again 1.2 eV excitation and comparing with the threshold values in Table II we see, that phonon energies in the range at least ~ 80 – 120 meV are required by Eq. (2) for enabling phonon-assisted absorption of the different divacancy configurations. While these energies are within the range of phonon energies of any of the SiC polytypes (the highest lattice-phonon energy is 120 meV), the phonon-assisted absorption with 1.2 eV (1030 nm) excitation is likely to be negligible even at room temperature (thermal energy $k_B T \approx 27$ meV, where $k_B$ is the Boltzmann constant and $T$ – the temperature), because the high-energy phonon modes are hardly populated. [We disregard here the temperature dependence of the bandgap, but if the band gap narrowing with increasing temperature is accounted for, this may lead to decrease of $\Delta E$ and, therefore, decrease of $\hbar\Omega$ in accord with Eq. (2).] However, in Ref. 15 a laser with wavelength of 976 nm (1.270 eV) is used. In this case, phonons of energies at least 10 – 50 meV are required for phonon-assisted absorption. Such phonons are readily available even at 150 K, quoted in [15] as the approximate limiting temperature above which the contrast between quenched and non-quenched PL is strongly reduced. Thus, not only the temperature, but also the excitation wavelength will have crucial importance for the observation of quenching at elevated temperatures. This statement, however, requires verification by means of further experimental work.

The described scenario suggests that the observation of quenching in the VV$^0$ PL excited with photon energies below ~ 1.3 eV requires generation of free electrons by this excitation. Since the photon energies below 1.3 eV (approximately) are not sufficient for photoionization of VV$^-$, the free electrons must be released from other defects. Let us consider first the n-SiC sample. The high N-donor concentration in this sample (in the $10^{17}$ cm$^{-3}$ range) together with the observation of the EPR signal of the carbon vacancy in the single negative charge state, (V$_C^-$) suggest that the charge-state distribution for most defects is dominated by the negative charge state(s). The higher charged states of V$_{Si}$, V$_{Si}^{2-}$ and V$_{Si}^{3-}$, as well as the double-negatively charged state of the N–V (N–V$^{2-}$) may also be populated, in accord with the energy diagram in Fig. 6. According to first-principles calculations,[26] N–V$^{2-}$ is quite shallow (cf. Fig. 6) and can be converted to single-negatively charged N–V$^-$ with low-energy photons. This may explain the strong luminescence associated with the negative charge state N–V$^-$ of the N–V defect (cf. Fig. 1). Thus, all above-mentioned negative charge states can emit electrons to the CB upon excitation below 1.3 eV photons, except for the divacancy (VV$^-$). Therefore, the latter accumulates in the negative charge and the luminescence from the neutral charge state quenches, as already discussed. We recall here that the EPR signal of the neutral N donors observable in this sample (n-SiC) serves as a monitor for the concentration of free electrons, and decays in synchrony with the quenching of the VV$^0$ PL as the free-electron population is depleting because of conversion of VV$^0$ to VV$^-$.

On the other hand, the EPR spectrum of the SI1 sample in dark displays contribution from the positively-charged carbon vacancy, V$_C^+$, but this does not preclude the existence also of the negatively charged states, V$_C^-$ and V$_C^{2-}$. In fact, the latter are associated with the so-called Z$_{1/2}$ centres,[32,33] known as efficient electron traps. The Z$_{1/2}$ centre (V$_C^{2-}$) has been observed by electrical measurements in the SI1 sample (not shown here), and it can be ionized by excitation of 1.2 eV, thus providing one possible source for free electrons. Thus, also in the case of the SI1 sample we attribute the quenching of the VV$^0$ PL to conversion to negative charge state.

Finally, the lack of quenching in the SI2 sample can be understood as due to lack of generation of free electrons by IR excitation of energy ~ 1.2 eV. Indeed, although the EPR spectrum in dark indicates the presence of the positively charged carbon vacancy, V$_C^+$, the negatively charged states (V$_C^-$ and V$_C^{2-}$) may not be present. This might be due to the presence of vanadium (visible in the PL spectrum, cf. Fig. 1), which contributes to compensating shallow donors and/or acceptors (vanadium is known to have amphoteric behaviour), thus pinning the Fermi level near the middle of the band gap. If the Fermi level is indeed pinned near the middle of the band gap, the divacancy may also be predominantly in its positively charged state VV$^+$. However, the application of 1.2 eV excitation will create and maintain certain quasi-equilibrium concentration of neutral divacancies, which produces the observed PL. We notice further that a repumping excitation with high enough photon energy may lead to generation of both electrons and holes, but such excitation opens also the path for conversion of VV$^-$ to VV$^0$, thus no quenching effect on the VV$^0$ PL is expected, neither below nor above 1.3 eV excitation.

VI. Dynamic model of the PL quenching

The preceding discussion explains qualitatively all experimental observations and suggests a simple model for the time dependence of the observed PL under quenching conditions. Since we are interested in reproducing with the model only the slow dynamics of accumulation of VV$^-$ for comparison with the experimental decays (Fig. 2), we disregard all fast processes leading to quick establishment of quasi-equilibrium concentrations of the various charge states in the initial stage of the IR excitation, except for the neutral and the negatively-charged states of the divacancy. By "quick establishment" we mean that the quasi-equilibrium concentrations of defects other than the divacancy and the traps generating electrons are established on a time scale faster than our time resolution of ~ 1 s. Thus, we disregard also the positively charged state VV$^+$, since it can quickly be converted to neutral in the beginning of the excitation, as already discussed. We assume further that free electrons are generated from a single type of electron traps by excitation to the CB with photons. When the photon energy is below the threshold for electron excitation from the negatively-charged state of the divacancy (quenching conditions), the latter can only capture electrons with certain probability and this capture is irreversible. The free-electron generation rate is proportional to the incident photon flux density and to the concentration of traps with electrons. On the other hand, free electrons can be captured back to empty traps or to neutral divacancies. The former process is reversible (by re-excitation with a photon) whereas the capture to divacancy is irreversible. Thus, one obtains the following system of coupled differential equations.

$$\frac{dN}{dt} = \sigma F N_T^- - \beta N N_T^0 - \gamma N N_{VV}^0, \tag{3}$$

$$\frac{dN_T^-}{dt} = -\sigma F N_T^- + \beta N N_T^0, \tag{4}$$

$$\frac{dN_{VV}^-}{dt} = \gamma N N_{VV}^0, \tag{5}$$

$$N_{VV}^0 + N_{VV}^- = N_{VV}, \text{ and} \tag{6}$$

$$N_T^0 + N_T^- = N_T. \tag{7}$$

If the excitation is switched on at the moment $t = 0$, we impose the following initial conditions:

$$N(0) = 0, \tag{8}$$

$$N_T^-(0) = N_T, \text{ and} \tag{9}$$

$$N_{VV}^0(0) = N_{VV}. \tag{10}$$

Here $N$, $N_T$ and $N_{VV}$ denote the free-electron concentration and the total concentrations of the traps and the divacancy, respectively. $N_{VV}^0$ and $N_{VV}^-$ denote the neutral and negatively-charged divacancy concentrations, which change dynamically, but their sum remains equal to $N_{VV}$. In a similar manner, $N_T^0$ and $N_T^-$ denote the concentrations of traps without and with electrons. Furthermore, $\beta$ denotes the electron-trapping coefficient for all the traps that have lost their electrons ($N_T^0$), and $\gamma$ is the corresponding coefficient for the neutral divacancies ($N_{VV}^0$). Finally, $\sigma$ denotes the absorption cross-section of the negatively-charged traps, and $F$ is the incident photon flux density (cm$^{-2}$s$^{-1}$). The photoluminescence intensity from VV$^0$ is proportional to its concentration $N_{VV}^0$.

The main limitation of this model is the assumption of only one type of "traps". Also, there is no account for possible spatial distribution of $F$ within the laser spot. Nevertheless, the model provides very convincing agreement with the experimental data for the SI1 sample, as illustrated in Fig. 2(a). The two fitting curves overlaying the experimental curves representing the decay of the PL4 line at two different excitation powers are calculated from the same model with all parameters fixed except for the photon flux density $F$, which differs by a factor of five in the two fitting curves. The five-fold increase of the excitation power (or $F$) mimics the experimental ~ 4.8-fold increase. The agreement between the experimental and the calculated decay curves is very good for this sample (SI1). However, no good fit could be obtained for the decay curves of the n-SiC sample displayed in Fig. 2(b), as illustrated further in the SM file.[28] Here we attribute the failure of the model to reproduce the shape of the decay curves for the n-SiC sample to the oversimplification of only one type of traps (described with a single electron-trapping coefficient $\beta$). Indeed, because of the higher N doping level, in addition to the carbon vacancy this sample contains several other defects in negative charge states, which can serve as traps (*vis.*, the N–V defect and the higher negative-charge states of $V_{Si}$).

The "scaling" property of the decay curves observed experimentally is nicely reproduced by the model. We notice, however, that this "scaling" property of the modelled decay curves only appears for certain relations between the parameters of the model. These relations may be informative in providing an estimation of the probabilities and cross-sections involved in the model, therefore we discuss them briefly below.

In order to solve numerically the system of Eqs. (3 – 7) with the given initial conditions (8 – 10), we bring the system to dimensionless form by dividing Eqs. (3 – 5) with $\beta N_T^2 = N_T / \tau$, where $\tau = (\beta N_T)^{-1}$ is some characteristic time. The rest of the equations are divided by $N_T$, and we introduce the dimensionless parameter $R_N = N_{VV} / N_T$ – the ratio of the total concentrations of divacancies and traps. In order to introduce the rest of the parameters, we write down one of the equations, Eq. (3), after division with $\beta N_T^2 = N_T / \tau$.

$$\frac{dn}{d\theta} = \Phi n_T^- - n n_T^0 - \delta n n_{VV}^0. \tag{11}$$

Here $\theta = t / \tau$ represents the dimensionless variable replacing the time variable, $\Phi = \sigma F / \beta N_T = \sigma F \tau$, $\delta = \gamma / \beta$, and all quantities denoted with lower case $n$ correspond to the quantities denoted in Eq. (3) with capital $N$ after normalization with $N_T$, i.e., $n = N / N_T$, $n_T^- = N_T^- / N_T$, etc.

In order to reproduce correctly the experimentally observed "scaling" property, the choice of the parameter $\delta = \gamma / \beta$ is not entirely arbitrary. We have observed that values of $\delta$ of the order of one (or larger) lead to decays which do not have the "scaling" property, therefore, the shape of the decay curves is not reproduced correctly. Thus, $\delta$ was decreased and the solutions at two different photon fluxes were compared until the value $\delta = 2 \cdot 10^{-3}$ was found to reproduce both the scaling property and the experimentally-observed decays, as shown in Fig. 2(a) (see also the SM file).[28] Note that $\delta$ is essentially the ratio between the rate of electron capture to the neutral divacancies and to the traps. Thus, the observation that $\delta \ll 1$ is equivalent to the assertion that the capture of free electrons to other traps is much more probable than capture to neutral divacancies. Such possible traps with large capture cross section are, for instance, the $Z_{1/2}$ centres, as well as the shallow donors. Thus, an electron experiences in average many cycles of photoionization and subsequent capture to the traps, before being finally captured to a divacancy, where it remains frozen in the absence of repumping excitation with energy enough for its photoionization.

Let us compare now our model for the quenching of the divacancy photoluminescence with the model of Ref. 15 presented in their Supplementary Information. While both models assume a single kind of traps, there are significant differences, both due to the different experimental conditions used in their work and ours, as well as conceptual. Thus, the model of [15] considers direct electron-hole pair generation, since above band gap excitation is also used in this work. We neglect the hole generation, assigning it to the "fast" processes which establish quasi-equilibrium hole concentration very shortly after the quenching has begun. Therefore, in our model only neutral and negatively charged divacancies are regarded; the positively-charged vacancies, as well as free holes may exist, but their concentrations establishes quickly after the ignition of continuous 1.2 eV excitation and do not change after. Thus, the separation of the ionization and recombination processes into "fast" and "slow" allows us to build up and deal with a much simpler model than that of Ref. 15. Furthermore, the model of [15] deals with capture cross sections for the carrier capture processes, whereas we prefer capture rates (defined by $\beta$ and $\gamma$ in Eqs. (3 – 5)). The photoionization processes are described by photoionization (or absorption) cross-sections in [15] and in our work. However, the most important conceptual difference between our model and Ref. 15 is the neglect of two-photon ionization processes in our model, which on the other hand is seen as the main reason for creating and developing the populations of the divacancy in its positive and negative charge states under the action of the probe pulse with photon energy suitable for quenching. We are quite confident that two-photon processes are negligible with our laser-power densities, which are about four order of magnitude lower than those used in Ref. 15. Finally, we notice that the charge-transfer levels taken in in [15] from Gordon *et al.*[37] differ somewhat from ours, presented in Table II and Fig. 6. In particular, their divacancy (0/−) level is taken at 1.1 eV below the conduction band, whereas ours is nearly at 1.3 eV below the CB. Thus, the 1.2 eV excitation falls below the excitation threshold in our model, but above it in the model of Ref. 15. Consequently, we attribute the quenching of the divacancy PL excited with < 1.3 eV photon energy entirely to the lack of the reversed process, photoionization of $VV^-$. In contrast, Ref. 15 attributes it to the different rates of creation of positively and negatively charged divacancies owing to two-photon processes. This is the main conceptual difference between our model and that of [15], but the observation of quenching at much lower power levels than those used in [15] seems to confirm our concept.

VII. Conclusions

We have studied the quenching properties of the divacancy-related PL in low excitation regime, with laser-power density at the sample at least four order of magnitude lower than using microscope-based setups in related work.[15,16] The experimental results are considered in the light of new more accurate *ab initio* results which provide new values for the charge transfer energy levels of the divacancies at the four different inequivalent configurations. Comparison between theory and experiment strongly supports the conclusion made in Ref. 15 that the dark state into which the neutral divacancies $VV^0$ are converted by excitation with photon energies below ~ 1.28 eV is the negative charge state VV$^-$. The PLE measurements made in this work allow precise determination of the energy thresholds associated with photoionization of VV$^-$ to VV$^0$ for each divacancy configuration and the experimentally-observed values are in very good agreement with the theoretical values for the (0/−) level of the four divacancy configurations. Additional confirmation for the conversion of VV$^0$ to VV$^-$ during quenching comes from the EPR measurements, as well as from the observed upconversion of the PL from the silicon-vacancy at elevated temperatures. Our low-excitation conditions as well as the physical model presented show that comprehensive understanding of the quenching phenomenon can be achieved by considering solely interaction (charge transfer) between the divacancies and other defects (traps). The dynamic model built on this notion is in excellent agreement with the experimental decays for one of the samples, hence, there is no need for involvement of two-photon processes which have been considered in Refs. [15,16]. Finally, we notice the existence of samples with strong VV PL but without any quenching effect (SI2 in our study), which can be understood in terms of lack of photo-generation of free electrons by photons with energies below ~ 1.28 eV.

*Acknowledgments*. Support from the Swedish Research Council (VR 2016-05362, VR 2016-04068), the Carl Trygger Stiftelse för Vetenskaplig Forskning (CTS 15:339), JSPS KAKENHI A 17H01056, the Swedish National Infrastructure for Computing (SNIC 2016/1-528) at the National Supercomputer Centre (NSC), Linköping University (LiU-2015-00017-60), and National Supercomputer Cluster in Hungary (Project gallium) is acknowledged. GA acknowledges the support from the National Research Development and Innovation Office of Hungary (NKFIH) within the Quantum Technology National Excellence Program (project no. 2017-1.2.1-NKP-21-2017-00001) and NKFIH grants Nos. 118161 and 127902.